\documentclass{rsauthor}
\usepackage{epsfig}
\usepackage{authblk}

\usepackage{graphicx,amsfonts,amssymb,amsmath,epstopdf,color,bm,amsthm}

\newcommand{\e}{{\epsilon}}

\newcommand{\brackets}[1]{\left( #1 \right)}

\newcommand{\R}{\mathbb{R}}

\renewcommand{\d}[1]{\, \, \mathrm{d} #1} % for integrals
\let\phi=\varphi
\newcommand{\abs}[1]{\left| #1 \right|} % for absolute value
\renewcommand{\v}[1]{\ensuremath{\mathbf{#1}}} % for vectors
\newcommand{\half}[0]{\tfrac{1}{2}}
\newcommand{\sech}{\operatorname{sech}}
\newcommand{\im}{\operatorname{Im}}
\newcommand{\sgn}{\operatorname{sgn}}
\newcommand{\uv}[1]{\ensuremath{\hat{\mathbf{#1}}}}
\newcommand{\beq}[0]{\begin{equation}}
\newcommand{\eeq}[0]{\end{equation}}
\newcommand{\pd}[2]{\frac{\partial #1}{\partial #2}} 
\newcommand{\pdd}[2]{\frac{\partial^2 #1}{\partial #2^2}}
\newcommand{\avg}[1]{\left< #1 \right>} % for average

\begin{document}
\title{Domain wall motion in magnetic nanowires: An asymptotic approach}
\author[1,2]{Arseni Goussev}
\author[3]{Ross G. Lund}
\author[3]{JM Robbins}
\author[3]{Valeriy Slastikov}
\author[3]{Charles Sonnenberg}
\affil[1]{Department of Mathematics and Information Sciences, Northumbria University, Newcastle upon Tyne, NE1 8ST, UK}
\affil[2]{Max Planck Institute for the Physics of Complex Systems, N{\"o}thnitzer Stra{\ss}e 38, D-01187 Dresden, Germany}
\affil[3]{School of Mathematics, University of Bristol, Bristol BS8 1TW, UK}
\label{firstpage}

\abstract{ We develop a systematic asymptotic description for domain wall motion in one-dimensional magnetic nanowires
  under the influence of small applied magnetic fields and currents and small material anisotropy.  The magnetization dynamics, as governed by the 
  Landau--Lifshitz--Gilbert equation,  is investigated via a
  perturbation expansion. We compute leading-order behaviour,
  propagation velocities, and first-order corrections of both
  travelling waves and oscillatory solutions, and find bifurcations between these two types of solutions. This treatment 
  provides a sound mathematical foundation for numerous results in the literature obtained through more 
  {\it ad hoc} arguments.}
  %puts  a number of results obtained through more ad hoc methods on a sound mathematical footing..}

\maketitle
\section{Introduction}   

The last two decades have witnessed a revolution in micromagnetics,
both in fundamental science and technology. It has long been
understood that ferromagnetic domains can be controlled through
external magnetic fields (Hubert and Sch\"{a}fer 1998). More recent is the discovery of a new
mechanism for domain wall dynamics through the interaction of
magnetization and spin-polarized currents (Berger 1978, 1984 and 1996; Slonczewski 1996). The ability to change (write) magnetic domains through the application of electrical
current has led to new designs of magnetic memory. One of the most
spectacular realizations is the so-called race-track memory (Parkin et al. 2008; Hayashi et al. 2008), a device with a fully three-dimensional geometry comprised
of an array of parallel nanowires on which domains (bits) may be read, transported, and written by applying currents.
% through the application of currents.

The dynamics of magnetic domain walls in ferromagnetic nanowires under
applied magnetic fields and electric current is one of the most
important problems in micromagnetics and spintronics. %Even though 
This
problem has been extensively studied experimentally (e.g.~Yamaguchi et al. 2004, Beach et al.~2005 and 2006;  Hayashi et al.~2007a and 2007b) and through
numerical simulations  (e.g. Hertel and Kirschner 2004; Weiser et al. 2010).

There have also been numerous theoretical studies of these phenomena based on the Landau-Lifshitz-Gilbert (LLG) equation.  The analysis is often simplified by taking the nanowire to be one dimensional, although even in one dimension, exact solutions of the LLG equation are available only in special cases (Schryer and Walker 1974; Goussev et al 2010).  A successful approach, introduced in this context by Schryer and Walker (1974) and generalised in Malozemoff and Slonczewski (1979), is to make the ansatz that 
under applied fields and currents, the static domain wall profile is preserved up to parameters describing its  position, orientation and scale.  The dynamics of these parameters is prescribed so as  to try to satisfy the LLG equation as nearly as possible. 

 This approach  has had numerous applications to domain wall motion in various regimes, and the results agree well with both numerical simulations and  experiments (see, e.g., Thiaville et al.~2005; Mougin et al.~2007; Bryan et al.~2008; Yang et al.~2008; Wang et al.~2009a and 2009b; Lu \& Wang 2010; Tretiakov \& Abanov 2010; Tretiakov et al.~2012).    However,  from a mathematical point of view, this approach is {\it ad hoc}, and difficult to justify;  the approximation is uncontrolled, and it is unclear how to obtain corrections to it.

%Thiaville 2005
%
%, and using some theoretical physics approaches (e.g. Schryer and Walker 1974; Mougin et al. 2007; Malozemoff and Slonczewski 1979),
%our current understanding of the process is far from complete.

Here we introduce a new approach to domain wall dynamics in thin nanowires.
We develop a systematic asymptotic expansion, regarding the applied field, material anisotropy and applied current as small parameters.  
% In this paper we investigate the dynamics of domain walls in
%thin nanowires by methods of applied mathematics and analysis. Using
%asymptotic analysis 
We derive formulas for two different types of solutions of the
LLG equations, namely travelling waves and oscillating
solutions, and obtain expressions
%. Under conditions of small material anisotropy, applied
%field and electric curren
%derive asymptotic formulas for the solutions and their those
%solutions and provide a description of
for their principal  characteristics, including velocity of propagation and frequency of oscillation.  In particular parameter regimes, these expressions agree with results obtained in previous studies, and put them on a sound mathematical foundation;  in this setting, they arise as solvability conditions for a system of inhomogeneous linear ODEs.  We  also obtain formulae for higher-order corrections, and carry out a systematic bifurcation analysis, including the values of applied fields and currents at which travelling-wave solutions break down, as well as bifurcations between one and two travelling waves, which arise through competition between the transverse magnetic field and material anisotropy.

%-----------------------------------------------------------------------------------------------------------

\subsection{Mathematical formulation}
The dynamics of magnetic domain walls (DWs) in thin ferromagnetic nanowires is governed by the Landau--Lifshitz--Gilbert (LLG) equations. We use the following dimensionless form of the equations (see Thiaville et al. 2005), which includes spin-transfer torque terms:
\beq
{\uv{m}}_t + \alpha \uv{m} \times {\uv{m}}_t=(1+\alpha^2)\uv{m}\times \v{H}(\uv{m})+J{\uv{m}}_x+\beta J{\bf \hat{m}}\times{\uv{m}}_x,
\label{LLG}
\eeq
where $\uv{m}(x,t)$ is the magnetization (a unit-vector field depending only on the coordinate along the wire $x$ and time $t$), $\alpha$ is the Gilbert damping parameter, $\beta$ the nonadiabatic spin transfer parameter, $\v{H}(\uv{m})$ the effective magnetic field and $J$ the applied current along the wire. 
The effective field, $\v{H}(\uv{m})$, is given by
\beq
\v{H}=A \uv{m}_{xx} + K_1 (\uv{m}\cdot \uv{e}_x) \uv{e}_x - K_2 (\uv{m}\cdot \uv{e}_y) \uv{e}_y + \v{H}_a,
\eeq 
where $A$ is the exchange constant, $K_1$ is the easy-axis anisotropy constant, $K_2 \geq 0$ is the hard-axis anisotropy constant, and $\v{H}_a=H_1 \uv{e}_x + H_2 \uv{e}_y + H_3 \uv{e}_z$ is a uniform applied field. Without loss of generality, for the rest of the paper we take $A=K_1=1$. Since we are interested in DW dynamics, we impose the following boundary conditions,
assuming that $\v{m}_t \to 0$ and $\v{m}_x \to 0$ as $|x| \to \infty$:
\beq \label{BC}
\v{m}(x) \times (K_1 (\uv{m}(x)\cdot \uv{e}_x) \uv{e}_x - K_2 (\uv{m}(x) \cdot \uv{e}_y) \uv{e}_y + \v{H}_a)  \to 0, \quad \text{ as } |x| \to \infty.
\eeq
%It is clear that, depending on parameters $K_2$ and $\v{H}_a$, equations \eqref{BC} may have multiple solutions. 
%NOTE: [JMR] Have commented out the preceding text.  Existence of multiple solutions is not ruled out by having tail to tail bc.s
%However, we consider only those supporting ``tail-to-tail'' domains.
{For definiteness, we consider boundary conditions supporting ``tail-to-tail'' domain walls, for which 
\begin{equation}\label{BC2}
\lim_{x\rightarrow -\infty} \uv{m}(x)\cdot\uv{e}_x < 0, \quad \lim_{x\rightarrow \infty} \uv{m}(x)\cdot\uv{e}_x > 0
\end{equation}
(``head-to-head'' domain walls may be treated similarly).}

Using polar angles $\theta(x,t)$ and $\phi(x,t)$, we express the magnetization as $\uv{m}=(\cos\theta, \sin\theta \cos \phi, \sin\theta \sin \phi)$ and represent \eqref{LLG} as a system of two PDEs 
\begin{align}
\alpha {\theta}_t +\sin\theta \phi_t &= (1+\alpha^2)F_1,  \label{LLG1} \\ 
-{\theta}_t+\alpha{\phi_t}\sin\theta &= (1+\alpha^2)F_2,  \label{LLG2}
\end{align} 
where $F_1$ and $F_2$ are given as
\begin{multline}
F_1=\theta_{xx}-\half(1+\phi_x^2)\sin 2\theta -\half K_2\cos^2\phi \sin 2 \theta +(1+\alpha^2)^{-1}\left(J\phi_x\sin\theta+\beta J\theta_x\right)\\- H_1\sin\theta +H_2\cos\theta \cos \phi + H_3 \cos\theta\sin\phi,
\end{multline}
\begin{multline}
F_2= \phi_{xx} \sin\theta + 2 \theta_x\phi_x \cos\theta + \half K_2\sin\theta\sin2\phi + (1+\alpha^2)^{-1}(\beta J\phi_x\sin\theta-J\theta_x) \\- H_2 \sin\phi+H_3\cos\phi.
\end{multline}
We %are interested in finding
look for solutions of equations \eqref{LLG}  depending on parameters $K_2$, $J$, $\v{H}_a$, $\alpha$, and $\beta$. While, in general, this problem may not be solvable with currently available PDE techniques, substantial progress can be made in understanding the {persistent} or long-time behaviour  by taking the some of the parameters -- namely  $\v{H}_a$, $K_2$ and $J$ -- to be small, and using the method of perturbation expansions.
%-----------------------------------------------------------------------------------------------------------

\section{Asymptotic analysis}
%**In this section we perform a formal asymptotic analysis of the LLG equations \eqref{LLG} assuming that $\v{H}_a$, $K_2$ and $J$ are \emph{small}.** 
We introduce  a parameter $\e \ll1$ and  set $\v{H}_a=\epsilon \v{h}_a = (\epsilon h_1, \epsilon h_2, \epsilon h_3)$, $K_2 = \epsilon k_2$, and $J=\epsilon j$. In order to capture the long-time behaviour of the problem \eqref{LLG},  we rescale time,  %we also change the time-scale of the problem \eqref{LLG} by 
defining $\tau=\epsilon t$.  It is straightforward to check that boundary conditions { \eqref{BC}--\eqref{BC2}} for $\uv{m}(x,t)$ become
\beq \label{BC1}
\uv{m} (\pm \infty, \tau) = (\pm 1, \e h_2, \e h_3) + O(\e^2)
\eeq
We seek solutions in the form of a regular perturbation expansion depending on $x$ and $\tau$ only:
\begin{align}
\theta(x,t)&=\theta_0(x,\tau) + \epsilon \theta_1(x,\tau) + \ldots,\\
\phi(x,t)&=\phi_0(x,\tau) + \epsilon \phi_1(x,\tau) + \ldots.
\end{align}
Substituting this expansion into the equations \eqref{LLG}, we obtain the following system of equations, to the leading order in $\epsilon$,
\begin{align}
\theta_{0,xx}-\half(1+\phi_{0,x}^2)\sin 2 \theta_0 &= 0,\label{LOS1} \\
\brackets{\phi_{0,x} \sin^2\theta_0}_x&=0. \label{LOS2}
\end{align}
The only physical { (finite micromagnetic energy)} solution of \eqref{LOS2} consistent with the boundary conditions \eqref{BC1} is $\phi_0(x,\tau)=\phi_0(\tau)$.  Equation \eqref{LOS1} then reduces to
\begin{equation}
\theta_{0,xx}=\half \sin2\theta_0,
\end{equation}
which (along with the restriction imposed by the boundary conditions) has a unique solution, given by the well known optimal DW profile
\beq
\theta_0(x,\tau)=2\arctan e^{-(x-x_*(\tau))}, \label{theta0-profile}
\eeq
as studied by Schryer and Walker (1974) and Goussev et al. (2010). Here $x_*(\tau)$ is a function representing the time-dependent position of the DW centre. Defining a moving coordinate $\xi=x-x_*(\tau)$, we see that $\theta_0(x,\tau)$ now depends on $x$ and $\tau$ only through $\xi$, which allows us to rewrite  our original perturbation expansion as
\begin{align}
\theta(x,t)&=\theta_0(\xi) + \epsilon \theta_1(\xi,\tau) + \ldots,\\
\phi(x,t)&=\phi_0(\tau) + \epsilon \phi_1(\xi,t) + \ldots.
\end{align}

We now focus on the functions $x_*(\tau)$ and $\phi_0(\tau)$, which,
to the leading order in $\epsilon$, describe the time-dependent
position and orientation of the DW, respectively. In what follows, we
use the prime ``$\; ' \;\,$'' to denote differentiation with respect
to $\xi$, and the dot ``$\; \dot{} \;\:$'' to denote differentiation
with respect to $\tau$.

%In the following we use the prime ``$'$'' and the dot ``$\dot{}$'' to denote differentiation with respect to $\xi$ and $\tau$, repectively. Inserting the revised ansatz into equations \eqref{LLG}, clearly we obtain the same solution for $\theta_0(\xi)$ from the leading order equation \eqref{LOS1}, but it tells us nothing about $x_*(\tau)$, and $\phi_0(\tau)$ . These functions represent the time-dependent orientation and position of the DW (to leading order) and will be the main dynamical quantities of interest. 

In order to determine the leading-order dynamics of the DW, we must
examine the LLG equations to the first order in $\epsilon$. To this
end, it is convenient to introduce new variables
\begin{align}\label{change_of_variables}
\Theta_1&:= \theta_1 -  (h_2 \cos \phi_0 + h_3 \sin \phi_0) \cos \theta_0, \nonumber\\
u&:=\phi_1 \sin\theta_0 +   h_2 \sin \phi_0 - h_3 \cos \phi_0,
\end{align}
which vanish at infinity due to the boundary conditions \eqref{BC1}. {(One way to motivate this transformation is through the M\"obius transformation which maps the boundary values for $\v{h}_a \neq 0$ into the boundary values for $\v{h}_a=0$.)} 
Investigating $\e$-order equations for $\theta_1$ and $\phi_1$, we obtain the following system of linear equations for $\Theta_1$ and $u$:
\begin{align}
L\Theta_1 &= f(\xi,\tau),\label{LinSysTheta}\\
Lu &= g(\xi,\tau)\label{LinSysu},
\end{align}
\noindent where $L$ is a self-adjoint linear differential operator defined as
\beq
L:=-\pdd{}{\xi}+\frac{\theta_0'''}{\theta_0'},
\label{L-definition}
\eeq
and $f$ and $g$ are given by
%\beq
\begin{align}
%f(\xi,\tau)
{ f} &={(1+\alpha^2)}^{-1}\sin\theta_0\left[-\alpha \dot{x}_* -\dot{\phi}_0-\beta j\right]- h_1\sin\theta_0-\half k_2\cos^2\phi_0 \sin 2 \theta_0,\\ 
%\eeq
%\begin{multline}
%g(\xi,\tau)
{ g}&={(1+\alpha^2)}^{-1}\sin\theta_0\left[ \dot{x}_* -\alpha\dot{\phi}_0 + j \right] +\half k_2\sin\theta_0\sin2\phi_0 \nonumber \\ & \qquad \qquad \qquad \qquad \qquad {} + (1- \cos 2\theta_0)( h_2 \sin\phi_0 +h_3\cos\phi_0).
%\end{multline}
\end{align}
A necessary condition for an equation of the form $Ly=F$, where $L$ is self-adjoint on $L^2(\R)$, to have a solution is that $F$ must be orthogonal to the kernel of $L$. The kernel of the operator $L$, defined by equation (\ref{L-definition}), consists of the single element $\theta_0'$. The required solvability conditions for the system \eqref{LinSysTheta}, \eqref{LinSysu} are then
\beq
\left<\theta_0',f\right>=0, \quad \left<\theta_0',g\right>=0 \label{SolCon},
\eeq
where angled brackets denote the $L^2$ inner product. Noting that $\theta_0'=-\sin\theta_0$ and $\theta_0''=\half\sin 2\theta_0$, it is straightforward to establish that $\left<\theta_0',\theta_0'\right>=2$, $\left<\theta_0',\theta_0''\right>=0$, $\left<\theta_0',\cos\theta_0\right>=0$ and $\left<\theta_0',1\right>=-\pi$. Using these results to evaluate the inner products in \eqref{SolCon}, we find  
\beq
\left<\theta_0',f\right>=2\left[h_1 + (1+\alpha^2)^{-1}(\alpha\dot{x}_* +\dot{\phi}_0+\beta j)\right]=0
\eeq
and
\begin{multline}
\left<\theta_0',g\right>= -k_2 \sin 2\phi_0 + \pi h_2 \sin\phi_0 - \pi h_3 \cos\phi_0 \\+ 2(1+\alpha^2)^{-1}(\alpha\dot{\phi}_0 - \dot{x}_*-j)=0.
\end{multline}
These equations lead to a system of two coupled ODEs for $\phi_0$ and $x_*$:
\begin{align}
\dot{\phi}_0 &= \mu(h_1,j) - \alpha \Gamma(\phi_0; h_2,h_3,k_2),\label{SolConPhi}\\
\dot{x}_*&=\nu(h_1,j) +\Gamma(\phi_0; h_2,h_3,k_2), \label{SolConx}
\end{align}
where 
\begin{align} 
\Gamma &= \frac{ \pi}{2}\left(h_2\sin\phi_0 - h_3 \cos \phi_0\right) - \frac{ k_2}{2}\sin2\phi_0, \label{Gamma} \\
\mu &=-h_1 -(1+\alpha^2)^{-1}(\beta-\alpha)j, \label{mu} \\
\nu &=-\alpha h_1-(1+\alpha^2)^{-1}(1+\alpha\beta)j .
\end{align}

Equations (\ref{SolConPhi}) and (\ref{SolConx}) determine the leading-order dynamics of the orientation and position of a DW. Considering $\v{h}_a$, $k_2$ and $j$ as parameters, we will use these equations extensively throughout the rest of the paper in order to characterize the behaviour of solutions in different parameter regimes.

%-----------------------------------------------------------------------------------------------------------

\subsection{First-Order Correction Terms}
The correction terms $\theta_1$ and $\phi_1$ can be computed by
solving equations \eqref{LinSysTheta} and \eqref{LinSysu}. The latter,
in view of the solvability conditions given by equations
(\ref{SolConPhi}) and (\ref{SolConx}), reduce to
\beq 
L\Theta_1=-\half k_2\cos^2\phi_0\sin2\theta_0,\eeq
\beq Lu=\left(h_2\sin\phi_0-h_3\cos\phi_0\right)\left({\tfrac{\pi}{2}}\sin\theta_0+\cos 2\theta_0-1\right).\eeq
We have already found one solution to the homogeneous problem $Ly=0$, given by $y_1=\theta_0'$. Now, taking into account that $\theta_0'=-\sech\xi$, we construct the Wronskian to obtain another { linearly independent} solution, $y_2= \xi\sech\xi+\sinh\xi$. (Note that $y_2$, unlike $y_1$, %does not vanish 
{diverges at infinity.)}  %Then, using these solutions, we construct a particular solution to the inhomogeneous problem $Ly=F$:
{ For $F$ square-integrable, the general square-integrable solution of the inhomogeneous equation $Ly=F$ is given by} % as follows:
\begin{equation}\label{green}
y(\xi) = \half\left( \int_0^\xi y_2(\eta)F(\eta)\, d\eta + a \right) y_1(\xi) - \half\left( \int_{-\infty}^\xi y_1(\eta)F(\eta)\, d\eta \right) y_2(\xi),
\end{equation}
%\begin{multline}
%{\color{blue} y(\xi)}  %y(\xi,\tau)
%=\int_0^{\xi}\half\sech\xi(\eta\sech\eta+\sinh\eta){\color{blue} F(\eta) }\d\eta\\-\int_0^{\xi}\half\sech\eta(\xi\sech\xi+\sinh\xi)%F(\eta,\tau)
%{\color{blue} F(\eta) } \d\eta, \label{green} \end{multline}
{where $a$ is an undetermined constant}.
%Validity of the above solution can be confirmed by a direct substitution. 
 Noting that $\sin\theta_0=\sech\xi$ and $\cos\theta_0=\tanh\xi$, it is straightforward to use (\ref{green}) to obtain
%\beq 
{
\begin{align} 
%\theta_1=-\half k_2\cos^2\varphi_0\xi\sech\xi+(h_2\cos\varphi_0+h_3\sin\varphi_0)\tanh\xi+a(\tau)\sech\xi,
\Theta_1&= -\half k_2\cos^2\varphi_0\xi\sech\xi+a(\tau)\sech\xi,\\
%\eeq
%\beq 
u&=(h_2\sin\phi_0-h_3\cos\phi_0)\tilde{u}+b(\tau)\sech\xi,
\end{align}
}
%\eeq
\noindent where 
\begin{multline}\tilde{u}=-\half\sech\xi\int_0^{\xi}\eta\sech\eta \d\eta+\half-{\tfrac{\pi}{4}}\sinh\xi\tanh\xi\\+\left(\xi\sech\xi+\sinh\xi\right)\arctan\left(\tanh\left(\tfrac{\xi}{2}\right)\right),\label{messy}\end{multline}
{and the functions $a(\tau)$ and $b(\tau)$ may be determined at second order in $\epsilon$.  (The integral term in (\ref{messy}) can be expressed in terms of polylogarithm functions, but the explicit closed-form expression is omitted).} 
%the result is somewhat cumbersome and thus not presented in this paper.
%-----------------------------------------------------------------------------------------------------------

\section{Domain wall dynamics }
We now turn our attention to solving the equations of motion \eqref{SolConPhi} and \eqref{SolConx} for the orientation, $\phi_0(\tau)$, and position, $x_*(\tau)$, of a DW. Firstly, from the form of the equations, it is simple to notice that the DW velocity $\dot{x}_*$ is given as a function of $\phi_0$ and the parameters $\v{h}_a,j$ and $k_2$. It is therefore sufficient to first consider only equation \eqref{SolConPhi} and solve it for $\phi_0(\tau)$. The DW position, $x_*(\tau)$, can then be found by integration of \eqref{SolConx}, given the result for $\phi_0(\tau)$.
For what follows, it is convenient to rewrite equation \eqref{SolConPhi} as
\beq \dot{\phi}_0 = F(\phi_0), \label{DynSys}\eeq
where the function $F$, defined as
\beq  F := \mu -\alpha\Gamma, \label{F-definition}\eeq
depends, parametrically, on $h_1$, $h_2$, $h_3$, $j$ and $k_2$. Here, $\Gamma$ and $\mu$ are defined in \eqref{Gamma} and \eqref{mu}, respectively.

\begin{figure}[h!]

\centering

\def\svgwidth{260pt}

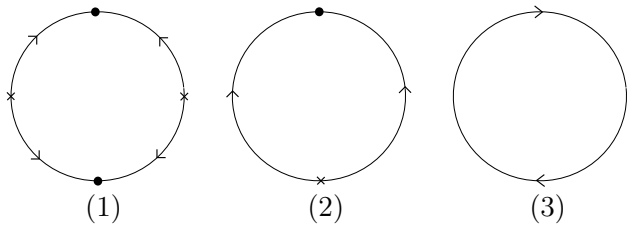

\caption{Schematic phase diagrams for $\phi_0$. Stable fixed points are indicated by a solid dot, and unstable ones with a cross. Case (1) corresponds to two stable Walker-like TWs, case (2) to a single stable TW, and case (3) to an OS.}

\label{phase}

\end{figure}

%We can now characterize the different solution types, as well as the bifurcations between them, by analyzing the zeros of $F$ and the points in the parameter space where the number of zeros change. 
%If $F(\phi_0)$ has real zeros, these correspond to fixed points $\phi_0^*$ of the equation, whose stability can be found by considering the sign of $F'(\phi_0^*)$.  
%If the parameter values are such that $F(\phi_0)$ has no zeros, then the system can be solved simply by separation of variables:
%\beq
%\frac{\d\phi_0}{F(\phi_0)} = \d \tau.
%\eeq
%We now examine the particular $F$ provided in \eqref{DynSys}, find the different possible solution types and then discuss each type in detail. For fixed parameters, the function $F$ is the sum of a constant, $\mu$, and a trigonometric polynomial of degree two, $\Gamma$, and is therefore periodic.  Let
%\begin{equation}\label{eq: mu_c}
%\mu_c = \mu_c(h_2,h_3,k_2) := \max_\phi  |\alpha \Gamma|,
%\end{equation}
%so that $F$ has no zeros for $|\mu| > \mu_c$.
%%{\color{blue} Let $\mu_c$ denote the smallest of $\mu$ such that $F$ has no zeros when $|\mu|>\mu_c$ (which will depend on $h_2,h_3$ and $k_2$), 
%There are three distinct possibilities for different types of solutions, represented graphically in Figure \ref{phase}:
We can now characterize the different solution types, as well as the bifurcations between them, by analyzing the zeros of $F$ and the points in the parameter space where the number of zeros change. 
Note that zeros of  $F$ correspond to the fixed points of Eq.~\eqref{DynSys}, whose stabilities are determined by the sign of $F'$ at the zeros.  
%If the parameter values are such that $F(\phi_0)$ has no zeros, then the system can be solved simply by separation of variables:
%\beq
%\frac{\d\phi_0}{F(\phi_0)} = \d \tau.
%\eeq
%We now examine the particular $F$ provided in \eqref{DynSys}, find the different possible solution types and then discuss each type in detail. For fixed parameters, the function $F$ is the sum of a constant, $\mu$, and a trigonometric polynomial of degree two, $\Gamma$, and is therefore periodic.  
There are three distinct possibilities (represented graphically in Figure \ref{phase}).
We
let
\begin{equation}\label{eq: mu_c}
\mu_c = \mu_c(h_2,h_3,k_2) := \max_\phi  |\alpha \Gamma|,
\end{equation}
(from Eq.~\eqref{F-definition}, $F$ has no zeros for $|\mu| > \mu_c$).
%{\color{blue} Let $\mu_c$ denote the smallest of $\mu$ such that $F$ has no zeros when $|\mu|>\mu_c$ (which will depend on $h_2,h_3$ and $k_2$), 
\begin{enumerate}
\item $F$ has 4 zeros. This occurs when $|\mu|<\mu_c$ and $k_2$ is dominant over the transverse field. Two of the zeros give stable fixed points, which correspond to two possible travelling wave (TW) solutions, with $\phi_0$ giving a fixed, constant orientation. Because this behaviour occurs when $k_2$ is dominant over $h_2$ and $h_3$, we refer to these TWs as ``Walker-like'' (called so after the case studied by Schryer and Walker (1974)).
\item $F$ has 2 zeros. This occurs when $|\mu|<\mu_c$ and  the transverse field is dominant over $k_2$. One of the zeros gives a stable fixed point, and hence a TW solution, which is different from the Walker-like TWs. 
\item $F$ has no zeros. This occurs when $|\mu|>\mu_c$. In this regime, we obtain an oscillating solution (OS), where the DW precesses with a time-dependent angular velocity, which is periodic, and propagates along the wire with a similarly periodic time-dependent velocity. While the precessional velocity always maintains its sign, the translational velocity oscillates between positive and negative values. Nevertheless, the average velocity is nonzero, giving rise to a mean drift of the DW.
\end{enumerate}

%-----------------------------------------------------------------------------------------------------------

\subsection{Travelling Waves}
As discussed above, when $F$ has real zeros, we obtain TW solutions. The fixed orientation angle of the DW can be found by solving $F(\phi_0)=0$. Using the substitution %\beq 
$z=\tan(\phi_0/2)$, 
%\label{subs}\eeq 
it is simple to show that $F=0$ is equivalent to the quartic equation
\begin{multline}p(z):=\left(\mu-\half \alpha\pi h_3\right)z^4+\left(-\alpha\pi h_2-2\alpha k_2\right)z^3+2\mu z^2\\+\left(-\alpha\pi h_2+2\alpha k_2\right)z+\mu+\half \alpha\pi h_3=0.\label{quartic}\end{multline}
The velocity of the TWs is constant, and can be found simply from the two equations (\ref{SolConPhi}) and (\ref{SolConx}). Substituting $\dot{\phi}_0=0$ in \eqref{SolConPhi}, we find
\beq 0= \mu-\alpha \Gamma, \quad \dot{x}_*= \nu +\Gamma, \eeq
and therefore
\beq \dot{x}_* =V_{TW}:=\nu + \frac{\mu}{\alpha}=- \brackets{\alpha +\frac{1}{\alpha}}h_1 - \frac{\beta}{\alpha}j. \label{TWVelocity} \eeq

This formula for the DW velocity is valid for any values of the parameters, so long as $|\mu|<\mu_c$. {Since} %Clearly 
$\mu$ depends linearly on $h_1$ and $j$, as does $V_{TW}$, %so 
the TW velocity can only increase up to a finite limit before the solution breaks down. This is analogous to the Walker breakdown.  { (Note, however, that the Walker solution satisfies the full nonlinear LLG equations, and the dependence of the velocity on the driving field is nonlinear, particularly near the breakdown field.)}

%-----------------------------------------------------------------------------------------------------------

\subsection{Oscillating Solutions}
When $|\mu|>\mu_c$, $F$ has no real roots, resulting in an OS. In this case, it is straightforward to separate variables in \eqref{DynSys}, obtaining
\beq
\tau = \int \frac{\d {\phi}_0}{F({\phi}_0)} =  \int \frac{2(1+{z}^2) }{p({z})}\d {z}, \label{tauofphi}
\eeq
%By substituting $z = \tan\phi_0/2$,  the integral above can be reexpressed as 
%\beq
%\tau = \int \frac{2(1+{z}^2) }{p({z})}\d {z}, \label{tauofx}
%\eeq
where $z = \tan \phi_0/2$ and $p({z})$ is the quartic polynomial given in Eq.~\eqref{quartic}.
{ While the integral \eqref{tauofphi} can be evaluated in terms of elliptic integrals, obtaining a general expression for $\phi_0(\tau)$ would be rather cumbersome.}
 %Evaluation of this integral, with all parameters being nonzero, would result in a very cumbersome expression for $\phi_0(\tau)$. 
Instead, we focus on computing the average precessional and translational velocities. (Note that in this section we are primarily concerned with general, average DW dynamics for arbitrary parameter values. We shall see in later sections that {in particular cases of interest,  it is not difficult to obtain explicit analytic expressions for $\phi_0(\tau)$.)}

Beginning with the average precessional velocity, we define
\beq \left\langle\dot{\phi}_0\right\rangle={\frac{2\pi}{T}},\eeq

\noindent where $T$ is the period of $\phi_0(\tau)$ and can be found from \eqref{tauofphi} as 
\beq T=\int_0^{2\pi}{\d \phi_0\over F(\phi_0)}.\eeq
In terms of $z$, the last expression is equivalent to
\beq T=2\int_{-\infty}^{\infty}{1+z^2\over p(z)}\d z.\eeq
{ Since we are working in the regime where $p(z)$ has no real roots, we let $a$ and $b$ denote its complex roots with $\im(a)>0$ and $\im(b)>0$, 
and write 
%we can factorize it as
\beq 
p(z)=\left(\mu-{\half\alpha\pi h_3}\right)(z-a)(z-\bar{a})(z-b)(z-\bar{b}).
\eeq
%where $a$ and $b$ are {\color{blue} complex roots of $p$. 
%  complex numbers with respective conjugates $\bar{a}$ and $\bar{b}$; these are the roots of $p(z)$. 
%Without loss of generality we can assume $\im(a)>0$ and $\im(b)>0$.   %and then 
We compute $T$ by contour integration} around a semicircular region in the positive half plane, yielding
\beq
T=\frac{4\pi i}{\mu-\half\alpha\pi h_3} \left({1+a^2\over(a-\bar{a})(a-b)(a-\bar{b})}+{1+b^2\over(b-a)(b-\bar{a})(b-\bar{b})}\right).
\eeq
The average precessional velocity is hence given by
\beq
\avg{\dot{\phi}_0} = \frac{\mu-\half\alpha\pi h_3}{2i} \left({1+a^2\over(a-\bar{a})(a-b)(a-\bar{b})}+{1+b^2\over(b-a)(b-\bar{a})(b-\bar{b})}\right)^{-1}.  \label{fullphiave}
\eeq

We now use this result to compute the average translational velocity $\avg{\dot{x}_*}$.
Recalling equations (\ref{SolConPhi}) and (\ref{SolConx}) we see that
\begin{align} 
\avg{\dot{\phi}_0}&=\mu-\alpha\avg{\Gamma},\label{avephi}\\
\avg{\dot{x}_*} &=\nu +\avg{\Gamma}.\label{avex}
\end{align}
Eliminating $\avg{\Gamma}$, we find the average velocity as
\beq
\avg{\dot{x}_*}=\nu + \frac{\mu-\avg{\dot{\phi}_0}}{\alpha} = V_{TW} -  \frac{\avg{\dot{\phi}_0}}{\alpha} \label{AverageOSVelocity}.
\eeq

%-----------------------------------------------------------------------------------------------------------

\subsection{Bifurcations}
We have classified three different solution behaviours (2 stable TWs, 1 stable TW, or OSs) and are now interested in finding { bifurcations} 
% the bifurcation points 
between these solution types. The bifurcation condition is straightforward. It represents the coincidence of two zeros of $F$ in the (five-dimensional) parameter space, and is given by the requirement that the equations
\beq
F(\phi_0) = 0 \quad \mathrm{and} \quad \pd {F}{\phi_0} = 0,
\eeq
are satisfied simultaneously. (Remember that $F$ depends, parametrically, on  $h_1$, $h_2$, $h_3$, $j$ and $k_2$.) With the substitution $z = \tan\phi_0/2$, this  problem is equivalent to that of finding coincident roots of two quartic polynomials, i.e.,
%\begin{multline}p(z):=\left(\mu-{\half \alpha\pi h_3}\right)z^4+\left(-\alpha\pi h_2-2\alpha k_2\right)z^3+2\mu z^2\\+\left(-\alpha\pi h_2+2\alpha k_2\right)z+\mu+{\half\alpha\pi h_3}=0,\label{BifCon1}\end{multline} 
%(same as equation \eqref{quartic}, and reproduced here only to enhance readability) corresponding to $F=0$, and
%\begin{multline} q(z):=\brackets{\half{\alpha\pi h_2}+\alpha k_2} z^4 + \brackets{-\alpha\pi h_3} z^3\\+(-6\alpha k_2) z^2 + (-\alpha \pi h_3) z + \alpha k_2 - \half{\alpha \pi h_2}=0, \label{BifCon2}
%\end{multline}
%corresponding to $F'=0$.
{
\begin{align}
p(z)&:=\left(\mu-{\half \alpha\pi h_3}\right)z^4+\left(-\alpha\pi h_2-2\alpha k_2\right)z^3+2\mu z^2\nonumber \\
&\qquad \qquad \qquad {}+\left(-\alpha\pi h_2+2\alpha k_2\right)z+\mu+{\half\alpha\pi h_3}=0,\label{BifCon1}\\
%(same as equation \eqref{quartic}, and reproduced here only to enhance readability) corresponding to $F=0$, and
 q(z)&:=\brackets{\half{\alpha\pi h_2}+\alpha k_2} z^4 + \brackets{-\alpha\pi h_3} z^3\nonumber\\
&\qquad \qquad \qquad {} +(-6\alpha k_2) z^2 + (-\alpha \pi h_3) z + \alpha k_2 - \half{\alpha \pi h_2}=0, \label{BifCon2}
\end{align}
corresponding to $F=0$ and $F'=0$ respectively.}

The locus of bifurcations can be found by computing the set of points in the parameter space where both of these equations are satisfied. Since there are four independent %variable 
parameters ($\mu$, $h_2$, $h_3$ and $k_2$) in the function $F$, the locus of bifurcations will be a hypersurface in four-dimensional space.
{Since the coefficients of $p$ and $q$ depend linearly on the four parameters, this hypersurface will be a ``cone-like'', i.e., invariant under overall rescaling.}

\begin{figure}[h!]

\centering

\def\svgwidth{180pt}

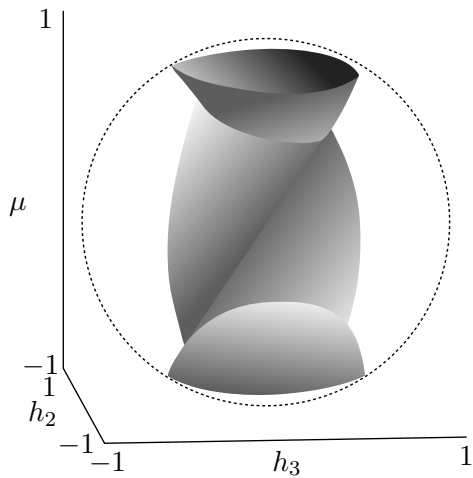

\caption{Three-dimensional sketch of bifurcation surface.}

\label{bif3d}

\end{figure}

The polynomials $p$ and $q$ have coincident roots if and only if their resultant vanishes. Calculating the resultant, and rearranging it to take form of a quartic polynomial in $\mu$, we obtain the bifurcation condition:
\begin{multline}\label{Res}
4096c^4 \, \mu^4 + 1024abc^3 \, \mu^3\\ + (16c^2(a^4+b^4)-1280c^4(a^2+b^2)+32a^2b^2c^2-2048c^6)\,\mu^2 \\ - 288abc^3(a^2+b^2+8c^2)\, \mu + 256c^8-192c^6(a^2+b^2)+48c^4(a^4+b^4-7a^2b^2)\\-4c^2(a^6+b^6+3a^4b^2+3a^2b^4) = 0,
\end{multline}
where $a=\alpha\pi h_2$, $b=\alpha\pi h_3$ and $c=\alpha k_2$.
This expression defines (implicitly) the bifurcation surface in the parameter space. In order to visualize and to develop intuition about the bifurcation surface, we now plot it by constraining the parameters with the condition $\mu^2 + h_2^2 + h_3^2+k_2^2=1$, and then substituting $k_2$, expressed in terms of $\mu$, $h_2$ and $h_3$, into equation \eqref{Res}. This gives an implicit surface in the three-dimensional $\mu, h_2, h_3$-space, which is contained in the unit ball; see Figure \ref{bif3d}. 

\begin{figure}[h!]

\begin{center}

\includegraphics[width=0.84\textwidth]{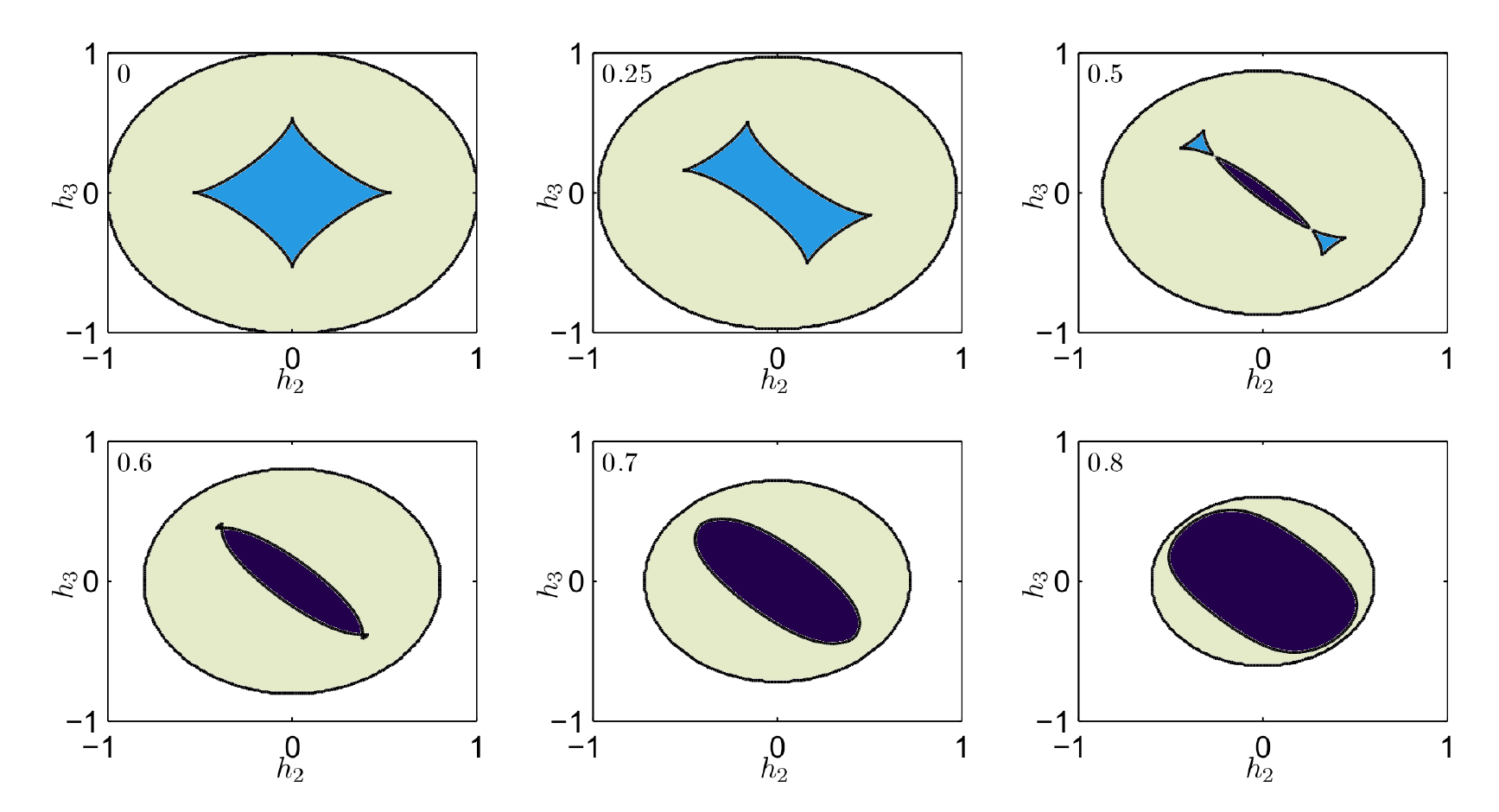}

\caption{(Colour online) Cross-sections of the bifurcation surface at constant positive values of 
the bifurcation parameter $\mu$ (shown in top left of panels). %The number in the top left of each panel is the corresponding value of $\mu$. 
The black regions correspond 
to the OS, while the light and grey regions (green and and blue online) correspond to 1 or 2 
stable TWs, respectively. The outer circles represent the cuts through 
the sphere $\mu^2 + h_2^2 + h_3^2=1$ (where $k_2=0$). The bifurcation 
surface cuts look identical for negative $\mu$, but are rotated by $\pi/2$.}

\label{bifslice}

\end{center}

\end{figure}

{The bifurcation surface in Figure \ref{bif3d} separates the parameter space into four distinct regions. 
There is a bounded ``Walker regime''  around the origin,  where the $k_2$-anisotropy dominates the dynamics (the origin is where $k_2=1$) and there are two stable Walker-like TW solutions. There are two  bowl-shaped surfaces which separate the Walker regime from the north and south poles of the unit ball (at $\mu = \pm 1$).  The OS is valid  in the interior of the bowls.  In the fourth region, which lies outside the bowls and the Walker regime, 
 the transverse field dominates the dynamics (the boundary of the ball is where $k_2=0$), and there is a single stable TW solution. Sections of the bifurcation surface at several constant values of $\mu$ are shown in Figure \ref{bifslice}).  }
% .  The interior of the bowls, which contain the poles $\mu = \pm 1$, .  Inside the bowls,  which contain the north anseparate the north and south polar regions (considering the ``poles'' to be at $\mu=\pm 1$) from a region around the equatorial line. The OS is valid in the polar regions, inside the bowls. Outside both the bowls and the closed central region is where there is one stable TW solution (the boundary of the ball is where $k_2=0$). 

Finally, we note that the roots $\mu$ of the quartic equation
(\ref{Res}) can be expressed explicitly in terms of values of the
parameters $k_2$, $h_2$ and $h_3$. However, these closed form analytic
expressions for the bifurcation points are rather lengthy
and, for this reason, are not reported in this paper.

%Finally, note that for fixed values of the parameters $k_2,h_2,h_3$, the equation to calculate the bifurcation points, as the driving parameter $\mu$ varies, is a quartic polynomial equation \eqref{Res}. Therefore the bifurcation points, and hence the critical values of $\mu$, can be calculated exactly. We don't provide these cumbersome expressions here.

%-----------------------------------------------------------------------------------------------------------

%%%%%%%%%%%%%%%%% SECTION 3: EXAMPLES %%%%%%%%%%%%%%%%%%%%%%%%%%%%%%%%%

\section{Examples}

%-----------------------------------------------------------------------------------------------------------

\subsection{Field-driven Motion $(j=0)$}

In the following section we set the applied current $j=0$, and examine the asymptotic behaviour of solutions in more detail, for several regimes of the parameters $h_1, h_2, h_3$ and $k_2$. This provides the leading-order behaviour of both travelling waves and oscillating solutions, and the bifurcation points between them.

%-----------------------------------------------------------------------------------------------------------

\subsubsection{Walker Case: $h_2=h_3=0$}

In the case of no transverse applied field $(h_2=h_3=0)$, an exact solution to the full nonlinear equation is known: the Walker solution. Here we show that our asymptotic method provides the essential features of the Walker solution, correctly predicts the so-called breakdown field, and also describes the motion beyond the breakdown field.

The ODE \eqref{DynSys} for $\phi_0$ reduces to 
\begin{equation}
\dot{ \phi}_0={\frac{\alpha k_2}{2}\sin 2\phi_0 - h_1}. \label{ODEWalker}
\end{equation}
In this case, there are just two possible solution types: $F(\phi_0)$ can now have either four zeros or no zeros (or two zeros at a bifurcation point). The critical driving field $h_{1,c}$ for the transition from the case of four zeros to that of none is easy to find; it occurs when
\beq
|h_1|=h_{1,c}:=\frac{\alpha k_2}{2}.
\eeq
This is the same as the breakdown field in the exact Walker solution, and is now the only bifurcation in the equation.

Firstly, consider the case $\abs{h_1} < h_{1,c}$. In this regime \eqref{ODEWalker} has four fixed points, and two of them are stable, corresponding to TW solutions. The orientation angle of the DW can be found as a function of $h_1$ and $k_2$ by solving
\begin{equation}
\sin 2 \phi_0 = \frac{h_1}{h_{1,c}} = \frac{2 h_1}{\alpha k_2},
\end{equation}
which again is the same as the value of $\phi_0$ found in the exact Walker solution. As expected for a TW solution, $\phi_0$ is found independent of $\tau$. The velocity of the travelling wave is given by the formula \eqref{TWVelocity} and reads
\begin{equation}
\dot{x}_*=-h_1\left(\alpha + \frac{1}{\alpha}\right).
\end{equation}
\noindent This is the leading-order term of the DW velocity predicted by the full nonlinear Walker solution. Combining the above expression for $\phi_0$ with the DW profile $\theta_0(\xi)$, given by equation (\ref{theta0-profile}), we obtain the full leading-order behaviour of the travelling wave solution corresponding to the Walker solution. In addition, our asymptotic analysis correctly reproduces an exact expression for the breakdown field $h_{1,c}$.

\begin{figure}[h]
\centering
\includegraphics[width=.8\textwidth, trim=20 0 20 10, clip=true]
{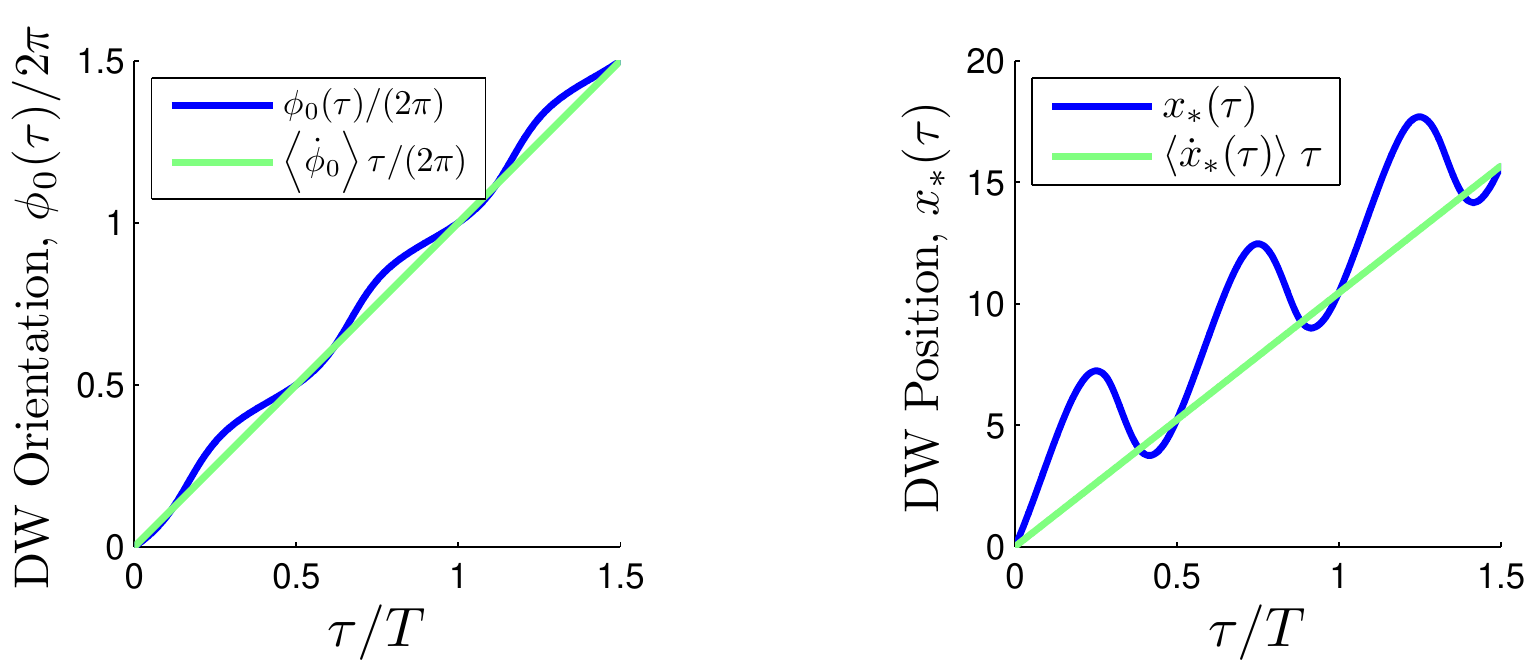}
\caption{(Colour online) Walker case: temporal profiles of $\phi_0(\tau)$ (left) and $x_*(\tau)$  (right). The dark curves (blue online) represent the explicit solutions, while the light lines (green online) show the average precession and propagation.}
\label{WC}
\end{figure}

Now consider the regime where $h_1>h_{1,c}$, in which the ODE
\eqref{ODEWalker} has no fixed points. The equation can be simply
solved by separation of variables to find an OS
\beq \label{OSS}
\tan \phi_0 = \frac{h_{1,c}}{h_1}-\frac{\sqrt{h_1^2 - h_{1,c}^2}}{h_1} \tan \brackets{\tau\sqrt{h_1^2 - h_{1,c}^2}}.
\eeq
In particular, we find that $\phi_0$ is a periodic (modulo $2\pi$)
function of $\tau$, with period
$T={2 \pi/}{\sqrt{h_1^2-h_{1,c}^2}}$. Physically, the magnetization DW
precesses with an angular velocity $\dot{\phi}_0$, which itself
oscillates in time (but remains close to a constant value, see
Figure~\ref{WC}). The average precessional velocity can be calculated
by integrating $\dot{\phi}_0$ over one period of the superimposed
oscillations:
\beq
\avg{\dot{\phi}_0} = \frac{\sqrt{h_1^2-h_{1,c}^2}}{2 \pi} \int_a^{a+T} \dot{\phi}_0 \d \tau.
\eeq
 Now, since
$$
\bigg[{\phi_0(\tau)}\bigg]_{\tau = a}^{\tau = a+T}= -\sgn(h_1) 2 \pi,
$$
we find
\begin{equation}
\avg{\dot{\phi}_0}= -\sgn(h_1)\sqrt{h_1^2-h_{1,c}^2},
\end{equation}
where $\sgn(\cdot)$ is the signum function. The precessional velocity changes monotonically with $h_1$, and approaches zero as $h_1 \rightarrow \pm h_{1,c}$. When $|h_1| \gg h_{1,c}$, the precessional velocity approaches $-h_1$. This is equivalent to setting $k_2=0$ in (\ref{ODEWalker}).

The average translational velocity can then be found from the formula \eqref{AverageOSVelocity}:
\begin{equation}
\avg{\dot{x}_*}=-h_1\left(\alpha+\frac{1}{\alpha}\right) + \frac{1}{\alpha}\sgn(h_1) \sqrt{h_1^2-h_{1,c}^2}.
\end{equation}
{This coincides with the travelling wave velocity for $|h_1| = h_{1,c}$, and decreases in magnitude as $|h_1|$ increases beyond $h_{1,c}$.
For $|h_1| \gg h_{1,c}$, the OS velocity approaches $-\alpha h_1$, which is
%Again, this reduces to the travelling wave velocity when $h_1 \rightarrow \pm h_{1,c}$, and approaches $-\alpha h_1$ when $|h_1| \gg h_{1,c}$. This limiting behaviour is 
consistent with the known exact solution of the LLG equations in the case of $k_2=0$ (the precessing solution -- see Goussev et al. 2010). 
%{\color{blue} Note that  the average speed of the OS is always less than the maximum TW speed $(1+\alpha^2)k_2/2  (unless $\alpha h_1 >  k_2$ is extremely large).   
Finally, we note that since we have an analytic expression for $\phi_0(\tau)$ in the oscillating regime, we can also find the velocity $\dot{x}_*(\tau)$ and hence the position $x_*(\tau)$ explicitly, though the expressions are quite unwieldy.
See Figure \ref{WC} for a plot of the DW position as a function of $\tau$. }

%-----------------------------------------------------------------------------------------------------------

\subsubsection{Maximal travelling wave velocity}

\begin{figure}[h]
\centering
\begin{minipage}{0.44\textwidth}
\includegraphics[width=\textwidth]{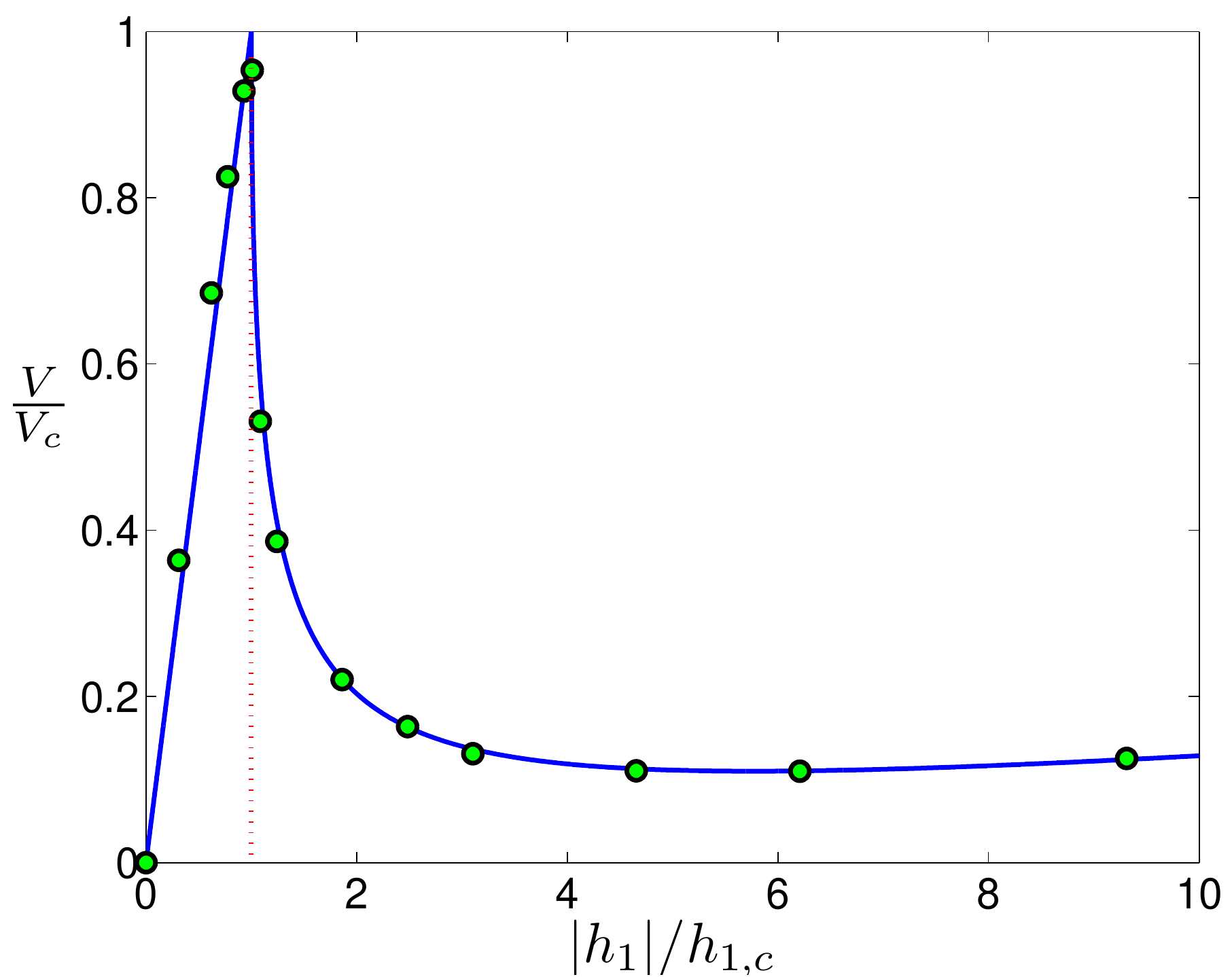}
\caption{Normalized average velocity of DW as a function of normalized driving field. Parameter values: $k_2=0.2, h_2=h_3=0.1.$ The curve is the analytical prediction; the points are  computed from numerically from the LLG equation \eqref{LLG}.}
\label{vvsH1}
\end{minipage}
\hfill
\begin{minipage}{0.44\textwidth}
\includegraphics[width=\textwidth]{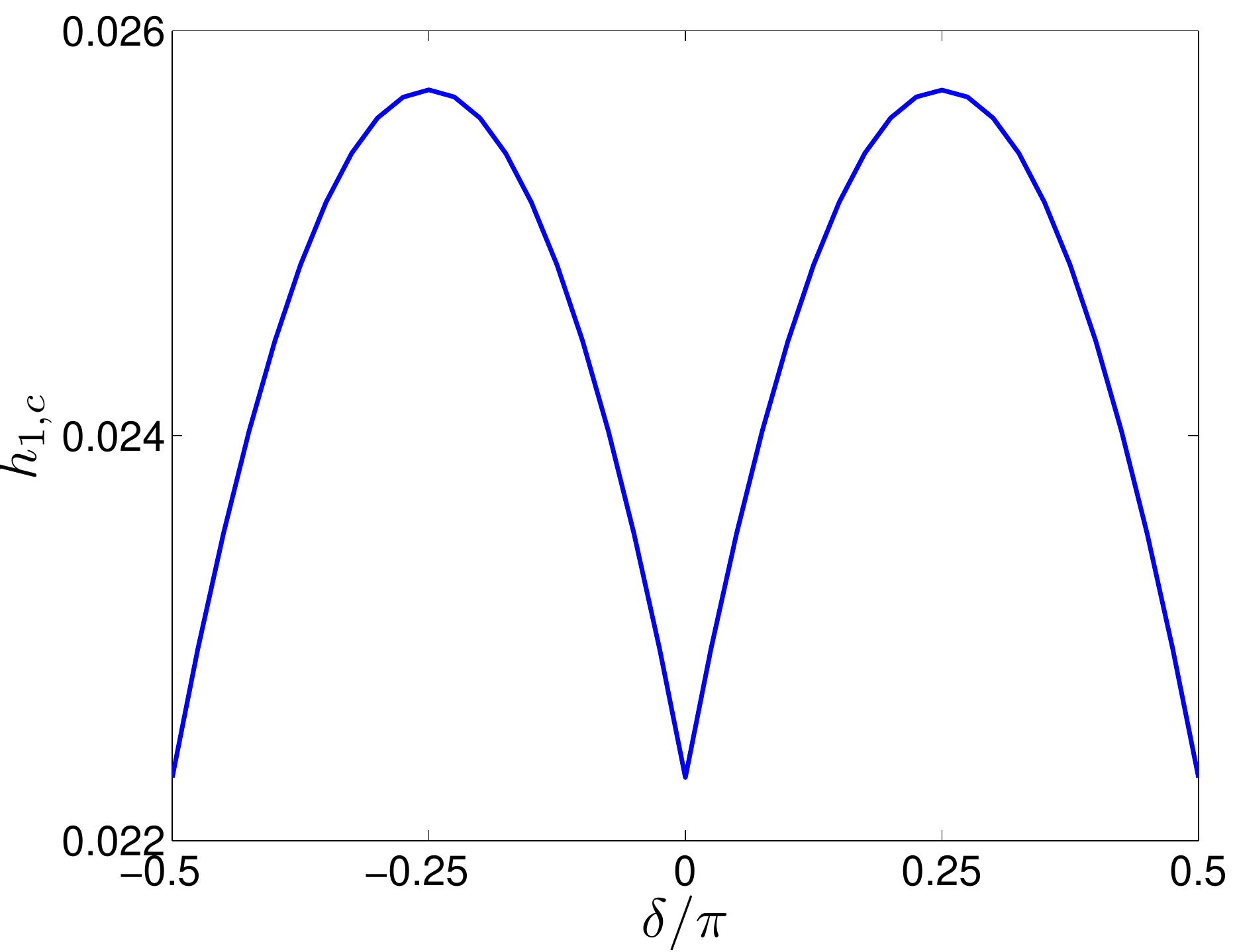}
\caption{Variation of critical driving field, $h_{1,c}$, with angle of transverse field, $\delta$. Fixed parameters: $(k_2=0.2, \alpha=0.1, h_T=0.1)$. %Note that changing $k_2$ or $h_T$ simply translates the curve up or down, and doesn't alter the horizontal positions of the maxima.
}
\label{Max_velocity}
\end{minipage}
\end{figure}

We investigate the maximal travelling wave velocity in the case of $j=0$. From \eqref{TWVelocity}, the TW velocity is given by%for the TW velocity
\beq
\dot{x}_*=-h_1(\alpha+\alpha^{-1}),
\eeq
which is valid %for any values of $k_2, h_2$ and $h_3$, so long as 
for $h_1\leq h_{1,c}$. 
The maximum velocity is obtained for $|h_1| = h_{1,c}$.
%n order to maximize the velocity, takes $h
%one simply needs to maximize $h_1$ without causing the TW to break down -- the problem of finding a critical field $h_{1,c}$.
The critical field $h_{1,c}$ is the largest root of %can be found by solving 
the quartic equation \eqref{Res}; % for  $h_1$,
% with %any 
equivalently (c.f.~Eqs.~\eqref{DynSys} and \eqref{Gamma}), we have that
%fixed values of $\alpha, k_2, h_2, h_3$, and taking the solution of largest magnitude. Equivalently, recalling that $\dot{\phi}_0 = -h_1 - \alpha \Gamma$ (cf Eq.~\eqref{DynSys}), we note that the travelling wave condition, $\dot\phi = 0$, implies that $h_1 = -\alpha \Gamma$. Therefore, the critical field $h_{1,c}$ is given by 
\beq \label{max phi}
|h_{1,c}|=\max_{\phi_0} |h_1| =\max_{\phi_0}\abs{\alpha\Gamma}.
\eeq
Figure~\ref{vvsH1} shows the dependence of the DW propagation velocity on $h_1$ for some representative nonzero values of $k_2, h_2$ and $h_3$, and, illustrates the sharp velocity maximum attained at $h_1 = h_{1,c}$.

%\noindent Recall now the equation for $\phi_0(\tau)$:
%\beq
%\dot{\phi}_0 = -h_1 - \alpha \Gamma.
%\eeq
%For a TW solution, $\phi_0$ is constant, and so $h_1 = -\alpha \Gamma$. Therefore, in order to calculate the critical field, we need to compute
%\beq
%|h_{1,c}|=\max_{\phi_0} |h_1| =\max_{\phi_0}\abs{\alpha\Gamma}
%\eeq
%for some fixed values of $k_2, h_2, h_3$. 

{Next,  letting $h_T=\sqrt{h_2^2+h_3^2}$ denote the strength of the transverse field and $\delta = \arctan(h_3/h_2)$ the angle between the transverse field and the hard-anisotropy axis, we seek to 
%writing we seek to
maximize $h_{1,c}$  with respect to $\delta$.  From Eq.~\eqref{max phi}, this is obtained by maximising $|\Gamma|$ over both $\phi_0$ and $\delta$.  Critical points of $\Gamma$ with respect to $\phi_0$ and $\delta$ are given by%THus given by critical points of $\Gamma$ The maximum values over the angle $\delta = \arctan(h_3/h_2)$ between the transverse field and the hard-anisotropy axis.  Writing
%Now, letting $h_T=\sqrt{h_2^2+h_3^2}$ and $\delta = \arctan(h_3/h_2)$, we maximize $|\Gamma|$ over both $\phi_0$ and $\delta$. In other words, we seek the  (\phi_0^c,\delta^c)$ of $\Gamma(\phi_0,\delta)$, as solutions of 
\begin{align}
{\partial \Gamma\over\partial\phi_0}&={\pi h_T\over2}\cos(\phi_0-\delta)-k_2\cos2\phi_0=0, \\
{\partial \Gamma\over\partial\delta}&=-{\pi h_T\over2}\cos(\phi_0-\delta)=0.
\end{align}
From these equations, it is straightforward to obtain
\beq \cos2\phi_0^c=0,\eeq
and hence $\phi_0^c \in \left\{{\pi\over4},{3\pi\over4},{5\pi\over4},{7\pi\over4}\right\}$.
Thus, we establish that $|\Gamma|$ (and hence $|h_{1,c}|$) is maximized when $\delta\in\left\{{\pi\over4},{3\pi\over4},{5\pi\over4},{7\pi\over4}\right\}$. This corresponds to $h_2=\pm h_3$. A numerical confirmation of this result is presented in Figure~\ref{Max_velocity}.}

%We also note that even the lowest points of the curve in
%Figure~\ref{Max_velocity}, which occur when either $h_2$ or $h_3$
%vanishes, correspond to a value of $|h_{1,c}|$ larger than the values
%of the critical field one would obtain in the case $k_2=0$ (with $h_T
%\not= 0$) or in the case $h_T = 0$ (with $k_2 \not= 0$). For example,
%with the parameter values used in the plot above, the smallest value
%of $h_{1,c}$ (say at $\delta=0$) is approximately 0.022, while
%$h_{1,c} \approx 0.016$ if $k_2=0$ and $h_T=0.1$, and $h_{1,c} \approx
%0.01$ if $h_T=0$ and $k_2=0.2$.

It is straightforward to show (we omit the argument) that the critical field is a strictly increasing function of $k_2$ and $h_T$. %, though this argument is omitted here. 
%thus, the critical field always increases with material anisotr
%Note that if $\delta$ is chosen from the set of maximal points, the critical field is given by
Choosing $\delta$ to maximise $h_{1,c}$, we obtain
\beq
h_{1,c}=\frac{\alpha k_2}{2} + \frac{\alpha \pi h_T}{2}.
\eeq
This analysis suggests that the most efficient way to get the fastest domain wall propagation, using constant applied fields, is to use a material with strong transverse anisotropy (large $k_2$) and apply a transverse magnetic field {at an angle $\pi/4$ with the anisotropy axis}. %Also, it is optimal to have the longitudinal field as strong as possible without causing the TW to break down.

%-----------------------------------------------------------------------------------------------------------

\subsection{Current-driven Motion $(\v{h}_a=0)$}

We consider the case of DW motion driven by a current applied along the wire. That is, we set $\v{h}_a=0$, and examine the DW dynamics under the influence of $j$ and $k_2$ only. {This problem has been widely studied in the literature (Thiaville et al. 2005; Mougin et al. 2007; Yan et al. 2010). It is straightforward to show that exact solutions of the LLG equation \eqref{LLG} can be found in two distinct regimes of $j$ and $k_2$. When $k_2=0$, there is a solution for arbitrary $j$ analogous to the precessing solution of field-driven motion {(Goussev et al. 2010)} given by
\beq
\theta(x,t)=\theta_0(x-x_*(t)), \quad \dot{x}_*(t)=\frac{-j(1+\alpha \beta)}{1+\alpha^2}, \quad \dot{\phi}(t)=\frac{\alpha-\beta}{1+\alpha^2}j,
\eeq
where $\theta_0$ is the profile given in equation \eqref{theta0-profile}.
When $k_2>0$ there is a TW solution for constant $j$ up to a certain critical value $j_W$, analogous to the Walker solution, given by
\beq
\theta=\theta_0\left(\frac{x-Vt}{\gamma}\right),\quad V = -\frac{\beta}{\alpha}j, \quad \sin 2\phi = \frac{j}{j_W}, \quad  j_W={\alpha(1+\alpha^2) k_2 \gamma\over2(\beta-\alpha)},
\eeq
 where $\gamma = (1+K_2\cos^2\phi)^{-1/2}$ is the usual Walker scaling factor.
This solution breaks down and produces an OS at values of $|j|$ above $|j_W|$. }
Using our perturbation analysis, we are able to characterize the leading-order behaviour of TWs and OSs, and the breakdown current. We then briefly discuss the propagation velocity of a DW in the current-driven case, where there are significant differences from the field-driven case depending on the relative sizes of the damping parameters $\alpha$ and $\beta$. 

\subsubsection{Leading-order behaviour}

Setting $\v{h}_a=0$ in equations (\ref{SolConPhi}) and (\ref{SolConx}) yields the equations for $\phi_0(\tau)$ and $x_*(\tau)$:
%\begin{align} 
\beq
 \dot{\phi}_0={\alpha k_2\over2}\sin 2\phi_0-\frac{(\beta-\alpha)}{1+\alpha^2}j, \quad % \label{phidotcurrent}\\
\dot{x}_*=-{k_2\over2}\sin2\phi_0-\frac{1+\alpha\beta}{1+\alpha^2}j. \label{xdotcurrent}
\eeq
%\end{align}
The critical current is straightforward to calculate from \eqref{xdotcurrent}, and is given by
\beq
 j_c:={\alpha(1+\alpha^2) k_2\over2(\beta-\alpha)}\label{criticalcurrent}.
\eeq
This determines the only bifurcation point in the system.
Note, unlike in the applied field (Walker) case, where the critical field matches the exact expression for the Walker breakdown field, the above expression for the critical current is not exact, but coincides with the leading-order term of the Taylor expansion  {in $k_2$} of the corresponding exact result. However, we shall see that an expression for the propagation velocity that we obtain \emph{is} exact (unlike in the applied field { case}).

We have two regimes to consider: $|j|<|j_c|$ and $|j|>|j_c|$.
%By the same techniques as before, 
{From the same arguments as in the previous section}, the case $|j|<|j_c|$ gives $\dot{\phi}_0=0$, requiring us to solve
\beq
{\alpha k_2\over2}\sin2\phi_0-\left(1+\alpha^2\right)^{-1}(\beta-\alpha)j=0.
\eeq
This yields two stable TW solutions (as in the Walker case of field-driven motion) which satisfy
\beq
\sin 2\phi_0=\frac{j}{j_c},\eeq
and have the propagation velocity given by
\beq\dot{x}_*=-{\beta \over\alpha}j .\eeq
This velocity is the same as that of the known exact TW solution.

\begin{figure}[h!]
\centering
\includegraphics[width=.53\columnwidth]{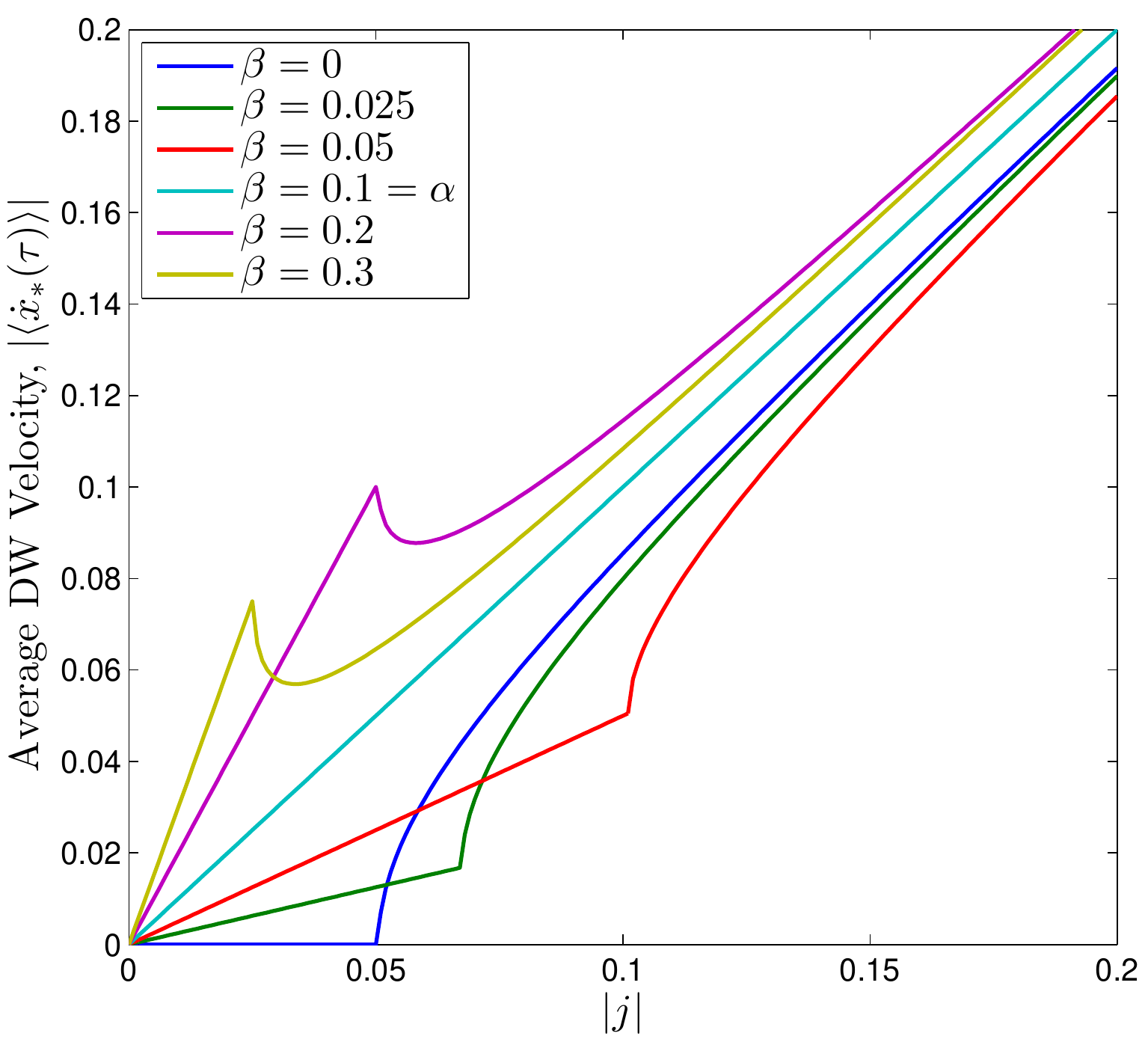}
\caption{(Colour online) Average velocity of DW as a function of applied current, for different values of the nonadiabatic damping parameter $\beta$. Fixed parameter values: $k_2=0.1, \alpha=0.1.$}
\label{vvsj}
\end{figure}

For the regime $|j|>|j_c|$, we get an oscillating solution similar to \eqref{OSS}, namely
\beq\tan\phi_0={j_c\over j}-{\sqrt{j^2-j_c^2}\over j}\tan\left({(\beta-\alpha)\over1+\alpha^2}\tau\sqrt{j^2-j_c^2}\right).\eeq
The corresponding average precessional and translational velocities are given by
\begin{align} 
\left\langle \dot{\phi_0}\right\rangle&=-\sgn(j)\left(\frac{\beta-\alpha}{1+\alpha^2}\right)\sqrt{j^2-j_c^2},
\\ 
\left\langle \dot{x}_*\right\rangle&=-\frac{\beta}{\alpha}j+\sgn(j)\left(\frac{\beta-\alpha}{\alpha(1+\alpha^2)}\right)\sqrt{j^2-j_c^2}.\label{current2}
\end{align}
It is interesting to note that in this current-driven case, the way the DW velocity varies with the applied current is subtly different from that in the applied field case. In particular, the current-velocity characteristic changes dramatically with the sign of $\beta - \alpha$ (see Figure \ref{vvsj}). 
When $\beta > \alpha$ the curves have a similar shape to the velocity-$h_1$ characteristic in Figure~\ref{vvsH1}; the velocity drops sharply as the current is increased above the critical current, then resumes a linear behaviour after that. However, { here} %a difference is that
the average velocity of the OS exceeds the maximal velocity of the TW (for fixed $\beta$) at an applied current not much greater than the critical current:
\beq
|j|>|j_c|\left(\frac{\beta^2(1+\alpha^2)^2 +(\beta-\alpha)^2}{\beta^2(1+\alpha^2)^2 -(\beta-\alpha)^2}\right) \sim |j_c|. 
\eeq
This is to be compared with the field-driven case, where the average velocity of the OS in the Walker regime only exceeds the maximum velocity of the TW solution when $h_1 \gg h_{1,c}$:
\beq
|h_1|>h_{1,c}\left(1+\frac{2}{\alpha^4+2\alpha^2}\right)\sim \frac{h_{1,c}}{\alpha^4}.
\eeq

 When $\beta=\alpha$ the critical current is infinite (this is clear from \eqref{criticalcurrent}), and so the TW behaviour persists for all applied currents. When $\beta < \alpha$ the velocity rises sharply above the critical current, and the average velocity of the OS always exceeds the maximal TW velocity. This behaviour is in accordance with that found in previous literature (Thiaville et al. 2005), and is particularly evident from the fact that when $\beta=0$ the TW velocity vanishes, and the DW does not undergo any motion until $j$ exceeds $j_c$, and the OS appears.  

%-----------------------------------------------------------------------------------------------------------

\subsubsection{Multiple Domain Walls}

\begin{figure}[h!]
\centering
\def\svgwidth{200pt}
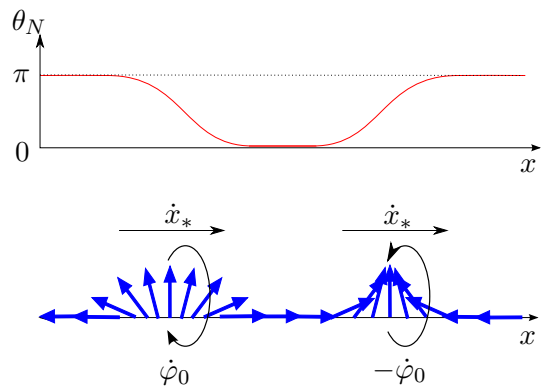
\caption{Adjacent DW profiles with leading-order translational and precessional velocities.}
\label{multipledws}
\end{figure}

%{In this section 
We examine the dynamics of a finite string of $N$ current-driven DWs. %g }
% and compare it with the field driven dynamics of a profile with multiple DWs. 
A typical multiple DW profile is given by
\beq \theta_N(\xi,\tau)=\displaystyle\sum_{n=1}^N\theta\left((-1)^{n+1}\xi_n,\tau\right),\eeq
where $\xi_n=x-x_{*,n}(\tau)$ and each $x_{*,n}$ is such that adjacent DWs do not overlap at $\tau=0$.
It is clear that when multiple DWs are present, a tail-to-tail profile must have head-to-head profiles on either side of it, and vice versa. Figure \ref{multipledws} shows schematically a tail-to-tail DW on the left followed by a head-to-head DW on the right, with their respective 
leading-order dynamics {(as calculated in the following analysis).} %One can imagine the profile sketched simply repeating as more DWs are added to the string.

 In Section II we {analysed} % looked at 
 the dynamics of the tail-to-tail profile {  $(\theta,\phi)$
 %$(\theta(\xi,\tau), %$, with corresponding azimuthal angle 
% $\phi(\xi,\tau))$, 
and found the %respective 
 leading-order precessional and translational velocities (\ref{SolConPhi}) and (\ref{SolConx}).  The dynamics of the head-to-head profile
 $(\tilde{\theta}, \tilde{\phi})$ % := \theta(-(x-\tilde{x}_*(\tau),\tau), \tilde{\phi})$
 may be similarly analysed, and one finds that   
 %the equations
 % ow we want to examine the head-to-head profile $\tilde{\theta}:=\theta(-(x-\tilde{x}_*(\tau)),\tau)$, with the corresponding azimuthal angle $\tilde{\phi}$. 
%
%For this profile, it is straightforward to apply the techniques from Section II to get  
%\begin{align}
%\dot{\tilde{\phi}}_0 &=\frac{(\beta-\alpha)}{1+\alpha^2}j+{\alpha k_2\over2}\sin2\tilde{\phi}_0 ,\label{SolConPhi2}\\
%\dot{\tilde{x}}_*&=-\frac{1+\alpha\beta}{1+\alpha^2}j+{k_2\over2}\sin2\tilde{\phi}_0. \label{SolConx2}
%\end{align}
%Comparison with Eqs~\eqref{phidotcurrent} and \eqref{xdotcurrent} reveals that $\tilde{\phi}_0=-\phi_0$, and therefore that
%By solving equation (\ref{SolConPhi2}) explicitly, we find that $\tilde{\phi}_0=-\phi_0$. Applying this result, we obtain}
%\begin{align}
\begin{equation} \dot{\tilde{\phi}}_0 =-\dot{\phi}_0 ,%\label{SolConPhi3}, 
\quad
\dot{\tilde{x}}_*=\dot{x}_*, \label{SolConx3} \end{equation}
%\end{align}
and thus $\dot{x}_{*,n}=\dot{x}_{*}$.
%\beq \dot{x}_{*,n}=\dot{x}_{*},\;\forall\;n\in\mathbb{N}.\eeq}
Unlike  the case of field-driven strings of DWs (Goussev et al. 2010), in which adjacent head-to-head and tail-to-tail DWs move in opposite directions but precess in the same direction, all current-driven DWs travel in the same direction, but with adjacent DWs precessing in opposite directions. }
%Nevertheless, this precessional property does not distort the overall multiple DW structure. 

\begin{figure}[h!]
\centering
\includegraphics[width=0.4\columnwidth]{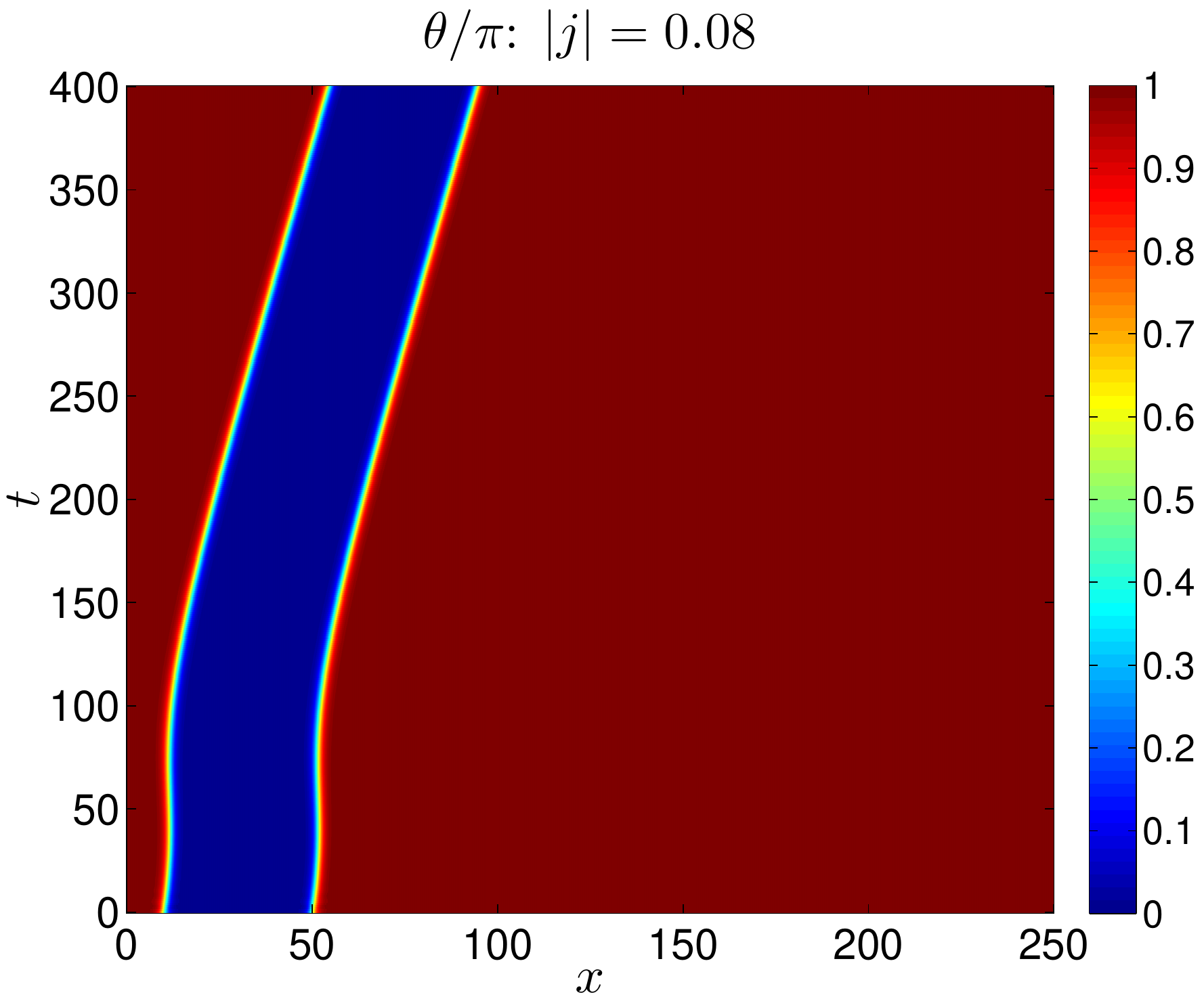}\hskip10pt
\includegraphics[width=0.4\columnwidth]{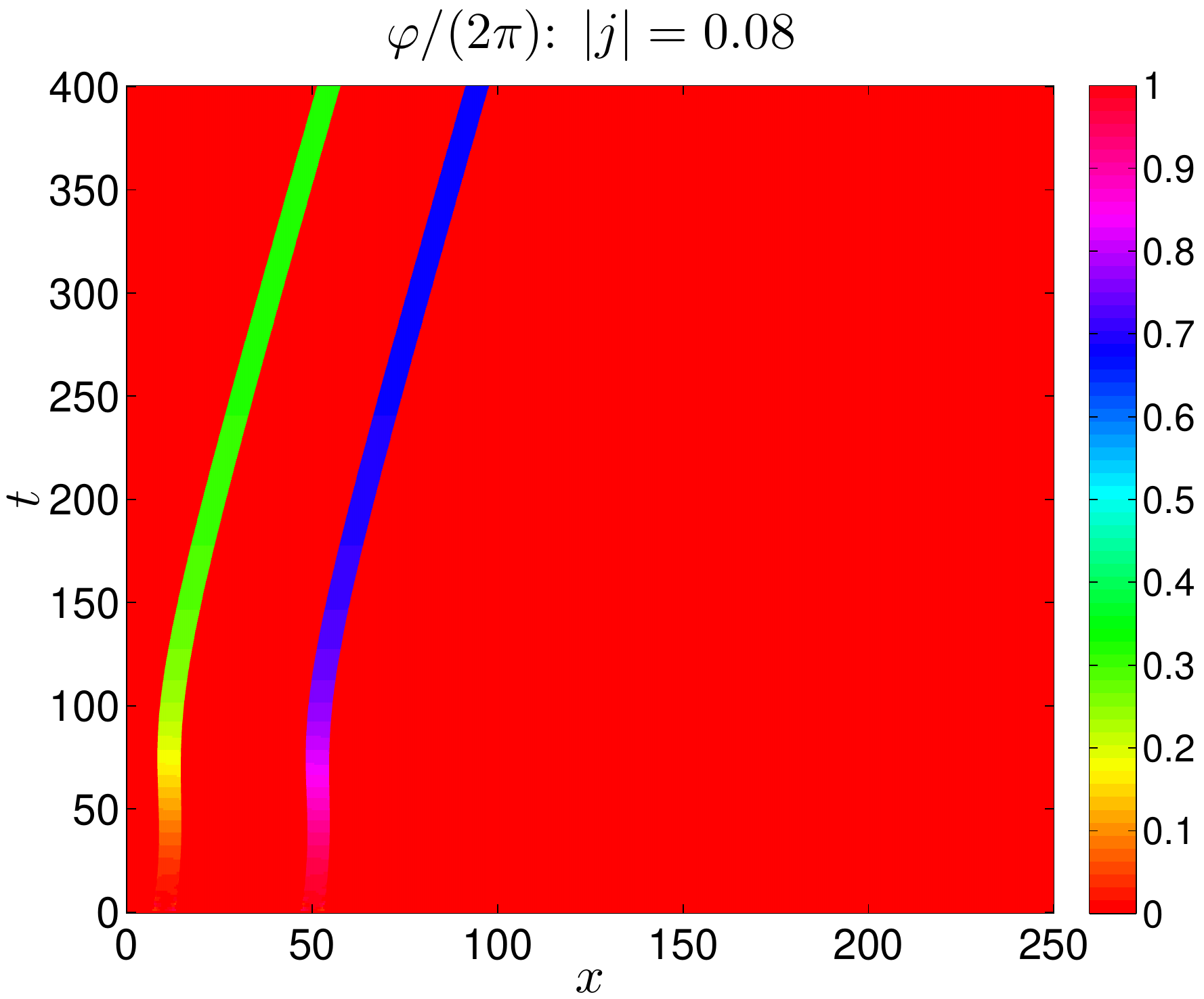}
\caption{(Colour online) Spatiotemporal Magnetization Profile---TW behaviour: two adjacent DWs driven by applied current $j=-0.08$. Fixed parameter values are $\alpha=0.1$, $\beta=0.2$, $k_2=0.2$.}
\label{currTW}
\end{figure}
\begin{figure}[h!]
\centering
\includegraphics[width=0.4\columnwidth]{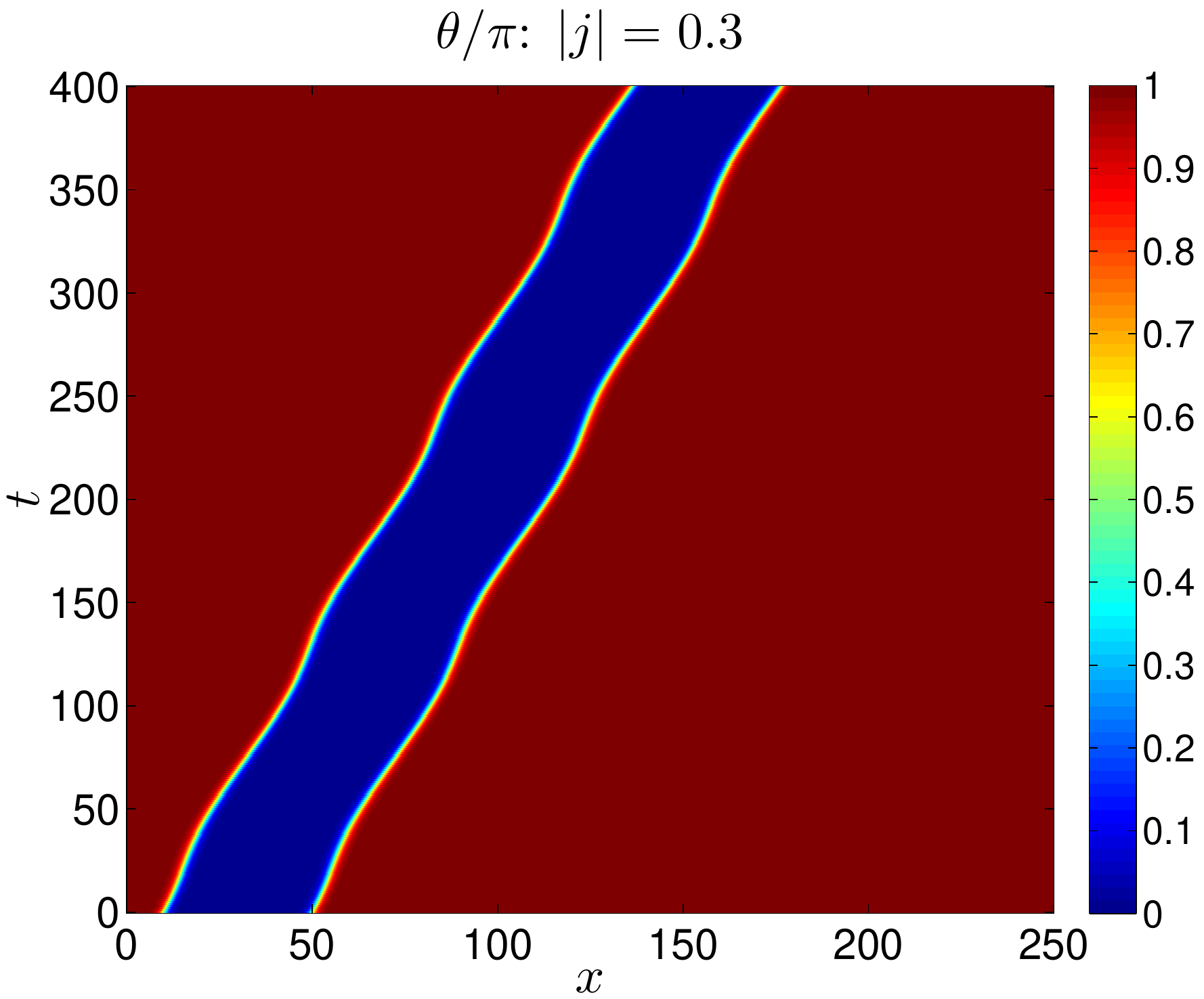}\hskip10pt
\includegraphics[width=0.4\columnwidth]{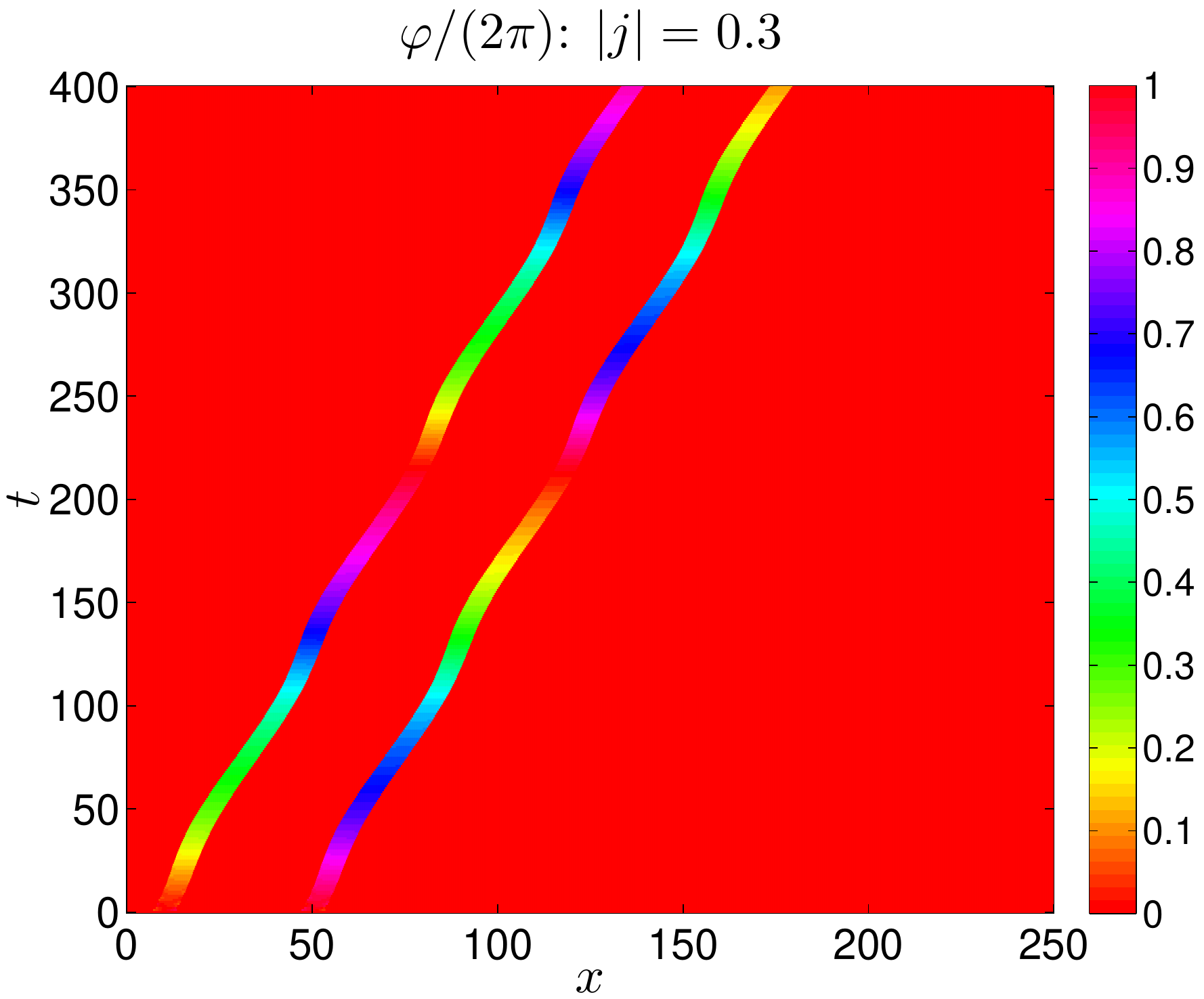}
\caption{(Colour online) Spatiotemporal Magnetization Profile---OS behaviour: two adjacent DWs driven by applied current $j=-0.3$. $\alpha=0.1$, $\beta=0.2$, $k_2=0.2$.}
\label{currOSC}
\end{figure}

Figures \ref{currTW} and \ref{currOSC} show the spatiotemporal magnetization profiles of a TW and an OS, obtained from numerical solution of the LLG equation  \eqref{LLG}. In each plot, the initial condition is given by two adjacent optimal DW profiles with orientation $\phi=0$. In the TW case we see that the translational velocities of the two DWs, after some initial transient, become constant and equal to each other. The DWs rotate in the opposite directions (at the same rate) and eventually, as expected, reach a steady state, in which the orientation of one DW is opposite to that of the other. In the OS case, we again find the DWs moving in the same direction, while rotating in the opposite directions (again at the same rate). The velocity of this OS is, in fact, greater than that of the TW shown. The ability to drive DWs in the same direction has a direct application in the recently proposed race-track memory architecture (Parkin et al. 2008; Hayashi et al. 2008)

%There, strings of oppositely magnetized domains separated by DWs are used to encode bits of information, and are driven along nanowires to and from reading and writing devices.

\section{Conclusion}
%In this  report 
 We have presented a new asymptotic analysis of domain wall dynamics in one-dimensional nanowires, governed by the Landau--Lifshitz--Gilbert equation. The approach is valid when the hard-axis anisotropy, applied magnetic fields and currents are small  compared to the exchange coefficient and easy-axis anisotropy.

Our approach covers both travelling-wave type and oscillatory domain wall motion. We present asymptotic formulas for the domain wall velocity and orientation angle in the travelling wave case, and for the average drift velocity and precession speed in the oscillatory case. Additionally, we are able to calculate the critical values of driving field (or current) for the transitions from travelling wave to oscillatory behaviour, as bifurcations in the associated dynamical system. This transition is analogous to the Walker breakdown, and is a generic feature of domain wall motion. The formulas given are valid for any combinations of the parameters (hard-axis anisotropy, applied fields and currents), provided the small parameter condition is met. We also present the leading-order domain wall magnetization profile and the next-order correction.

We discuss several special cases  and demonstrate that our approach  accurately reproduces the essential features of domain wall motion. We compare results of our asymptotic analysis with both numerical and exact analytical results (when available) and find them to be in good agreement.

\ack{We would like to acknowledge support from the EPSRC (grant EP/I028714/1 to VS, and doctoral training awards to CS and RGL).}

\end{document}